\documentclass[11pt,twoside]{article}

\usepackage{asp2006}

\begin{document}
\title{Infrared Emission from High-Redshift Galaxies in Cosmological SPH Simulations}   %%% Fill in title
\author{Kentaro Nagamine, Tae Song Lee, and Jun-Hwan Choi}   %%% Fill in author names
\affil{Department of Physics and Astronomy, University of Nevada Las Vegas, 4505 S. Maryland Pkwy, Box 454002, Las Vegas, NV 89154-4002, U.S.A. }    %%% Fill in author affiliations

\begin{abstract} %%% Abstract to run on from here.
We compute the infrared (IR) emission from high-redshift galaxies in 
cosmological smoothed particle hydrodynamics (SPH) simulations 
by coupling the output of the simulation with the population 
synthesis code `GRASIL' by Silva et al. 
Based on the stellar mass, metallicity and formation time
of each star particle, we estimate the full spectral energy 
distribution (SED) of each star particle from ultraviolet (UV) to IR, 
and compute the luminosity function of simulated galaxies in 
the {\it Spitzer} broadband filters for direct comparison with 
the available {\it Spitzer} observations. 
\end{abstract}

%%% MAIN BODY OF TEXT GOES HERE. CONSULT "INSTRUCTIONS FOR AUTHORS USING
%%% LATEX2E MARKUP", SECTIONS 2.3-2.6 FOR HELP WITH EQUATIONS, FIGURES,
%%% AND TABLES.

\section{Introduction}   

A significant fraction of the energy density in the Universe comes from 
the IR emission reprocessed by dust.  Therefore we need to take
into account this conversion process from stellar UV photons to 
dust thermal emission in IR in order to estimate the 
{\it total} cosmic SFR density in the Universe properly.  
The observations by the {\it Spitzer Space Telescope} give us 
excellent opportunities to test our models of galaxy formation and evolution. 

Theorists have utilized both semianalytic models and cosmological 
hydrodynamic simulations of galaxy formation to understand the history 
of cosmic star formation \citep[e.g.,][]{Kau00, Nag00, Nag01b}.  
However, so far the modeling of dust extinction and the energy conversion 
from UV to IR have not been considered in detail in these modeling 
\citep[but see][]{Lacey07, Lacey10}. 
In this article, we report the preliminary results of our attempt to 
compute the IR emission from simulated galaxies in cosmological hydrodynamic 
simulations. 

%%%%%%%%%%%%%%%%%%%%%%%%%%%%%%%%%%%%%%%%%%%%%%%%%%

\section{Numerical Simulation and Method}

%\vspace{0.3cm}
%\noindent
%{\bf Numerical Simulation and Methods:}
Cosmological hydrodynamic simulations allow us to model galaxy formation starting from high-$z$ to the present time without any assumptions on the dynamics of collapsing gas or merging dark matter halos.  
We use the updated version of the {\small GADGET-3} SPH code \citep[originally described in][]{Springel05e}.
Our code includes radiative cooling by H, He, and metals \citep{Choi09c}, heating by a uniform UV background radiation, 
star formation \citep[SF,][]{Choi09a}, supernova feedback, a phenomenological model for galactic winds \citep{Choi10}, and a sub-resolution model of multiphase ISM \citep{Springel03b}.
Once the gas density exceeds the SF threshold density, a star particle is allowed to be born from a gas particle at each time-step in order to statistically reproduce the computed SFR on average. 
We use a series of simulation runs with different box sizes (comoving 10, 34, and 100$h^{-1}$Mpc) and resolution to cover a wide range of galaxy masses.

\begin{figure}
\epsscale{0.69}
\plotone{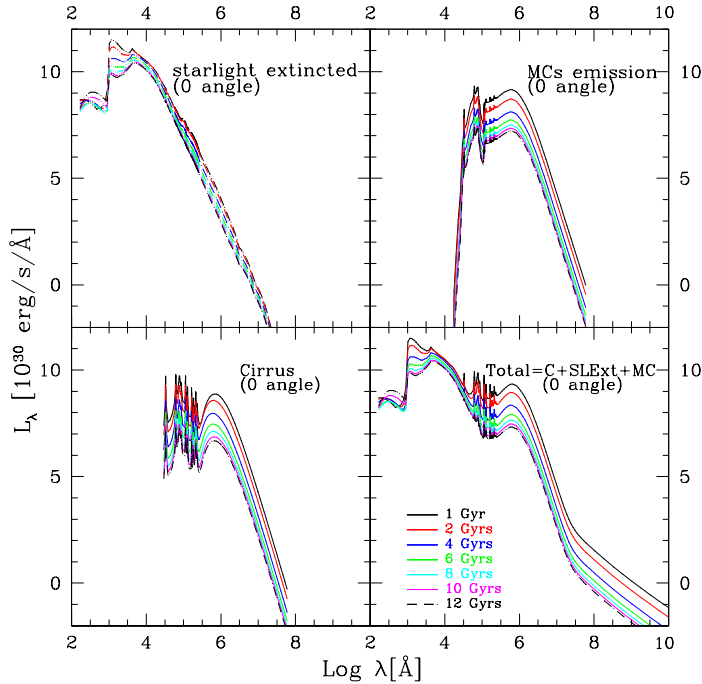}
\caption{SEDs of different emitting physical components within the GRASIL model: stars (top left), molecular clouds (MC; top right), cirrus (bottom left), and total (bottom right). The time evolution of a single stellar population is shown from $t=1$\,Gyr to 12\,Gyr from top to bottom in each panel. Dust absorbs the UV photons, and re-emit the energy in the IR as the molecular cloud and cirrus emission.  }
\label{fig:sed_compo}
\end{figure}

The GRASIL code \citep{Silva98} computes the time-dependent SEDs of simulated galaxies in the wavelengths of $100\AA < \lambda < 1$\,m for a given stellar population (Figure\,\ref{fig:sed_compo}), taking into account of the dust extinction processes that convert the UV photons into IR. We apply the GRASIL code to compute the IR emission of each star particle based on its mass, formation time, and metallicity, treating each star particle as an instantaneous starburst of single stellar population.   We then co-add the spectra of all constituent star particles to obtain the spectra of simulated galaxies. The \citet{Salpeter55} IMF has been assumed. 

%%%%%%%%%%%%%%%%%%%%%%%%%%%%%%%%%%%%%%%%%%%%%%%%%%

\vspace{-0.5cm}
\section{Results}

\begin{figure}
\epsscale{0.7}
\plottwo{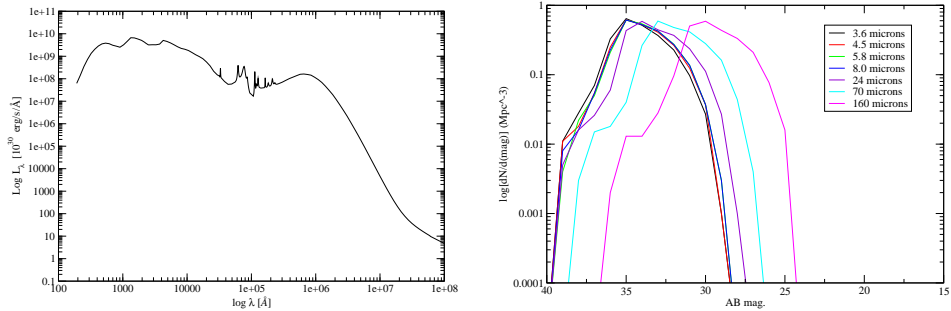}{nagamine_k_3.eps}
\caption{{\it Left:} Total SED from UV to IR of a simulated galaxy at $z=3$, 
computed by coupling the GADGET-3 output and the GRASIL code. 
{\it Right:} AB magnitude luminosity functions (LFs) in the {\it Spitzer} broadband filters of 160, 70, 24, 8, 5.8, 4.5, 3.6\,$\mu$m, from right to left. The latter four LFs are almost overlapping each other. 
}
\label{fig:sed_sim}
\end{figure}

The left panel of Figure\,\ref{fig:sed_sim} shows an example SED of a simulated galaxy at $z=3$ with a stellar mass $M_\star \sim 10^{10} M_\odot$.  One can see that some UV photons are absorbed by dust and re-emitted as the dust thermal emission with a peak near 100\,$\mu$m. 

Once we obtain the full SED of simulated galaxies, we first compute the AB magnitudes using the {\it Spitzer} broadband filters, and examine the luminosity functions as shown in the right panel of Figure\,\ref{fig:sed_sim}. We find that the simulated galaxies are brighter in the longer wavelength filters, although we still need to identify the exact cause of this trend. 

\begin{figure}
\plottwo{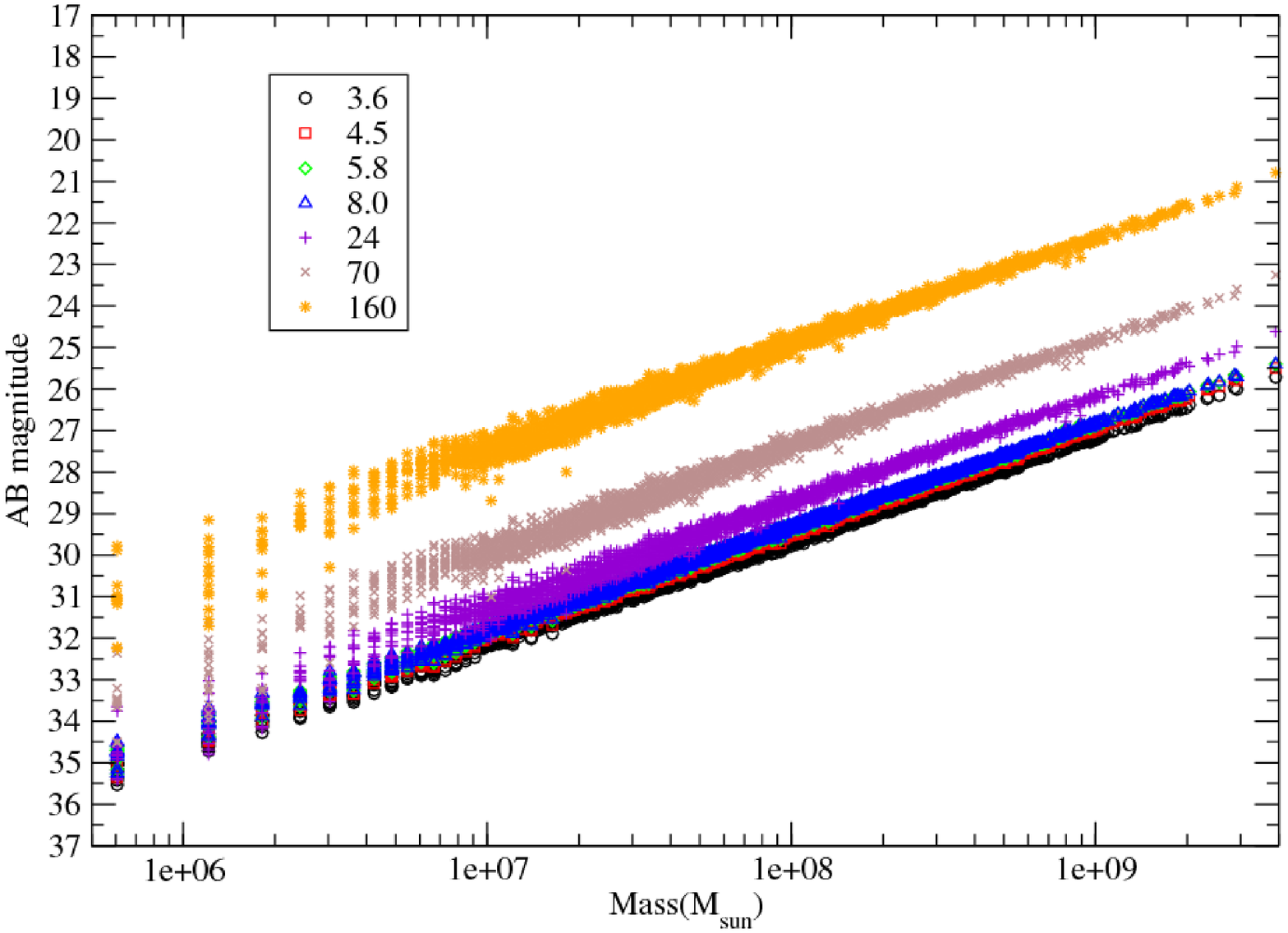}{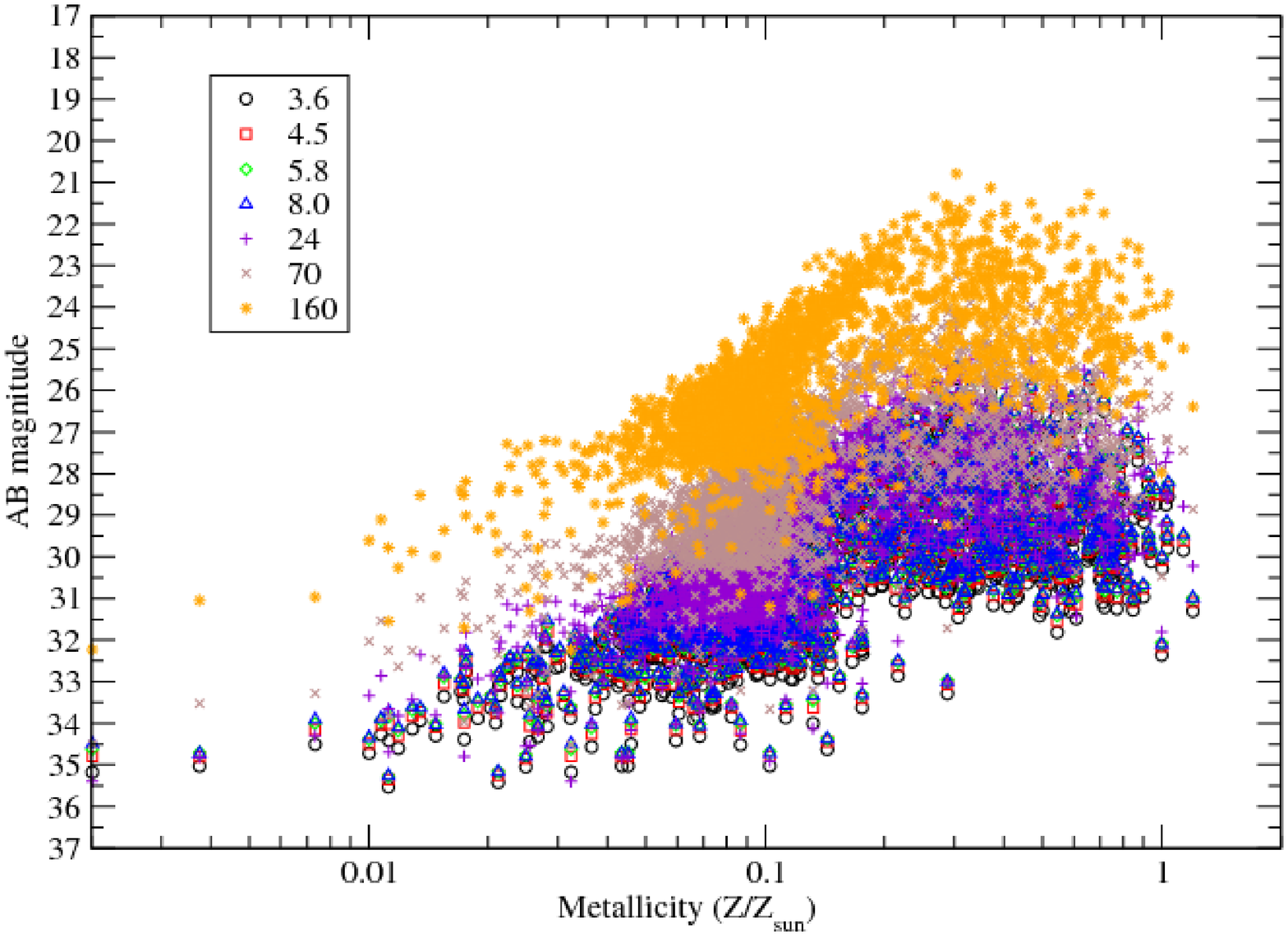}
\caption{Stellar mass vs. AB magnitude ({\it left}) and galaxy metallicity vs. AB magnitude ({\it right}) for simulated galaxies at $z=3$. }
\label{fig:mag}
\end{figure}

Figure\,\ref{fig:mag} shows the relation between the AB magnitude, stellar mass, and metallicity of simulated galaxies at $z=3$. As expected, more massive galaxies have brighter AB magnitudes.  We also see that in general, the higher mass galaxies have higher metallicities. However, the scatter in metallicity for the luminous galaxies is much larger compared to when it is plotted against galaxy stellar mass, presumably owing to dust extinction and re-emission processes. 

\begin{figure}
\epsscale{0.7}
\plottwo{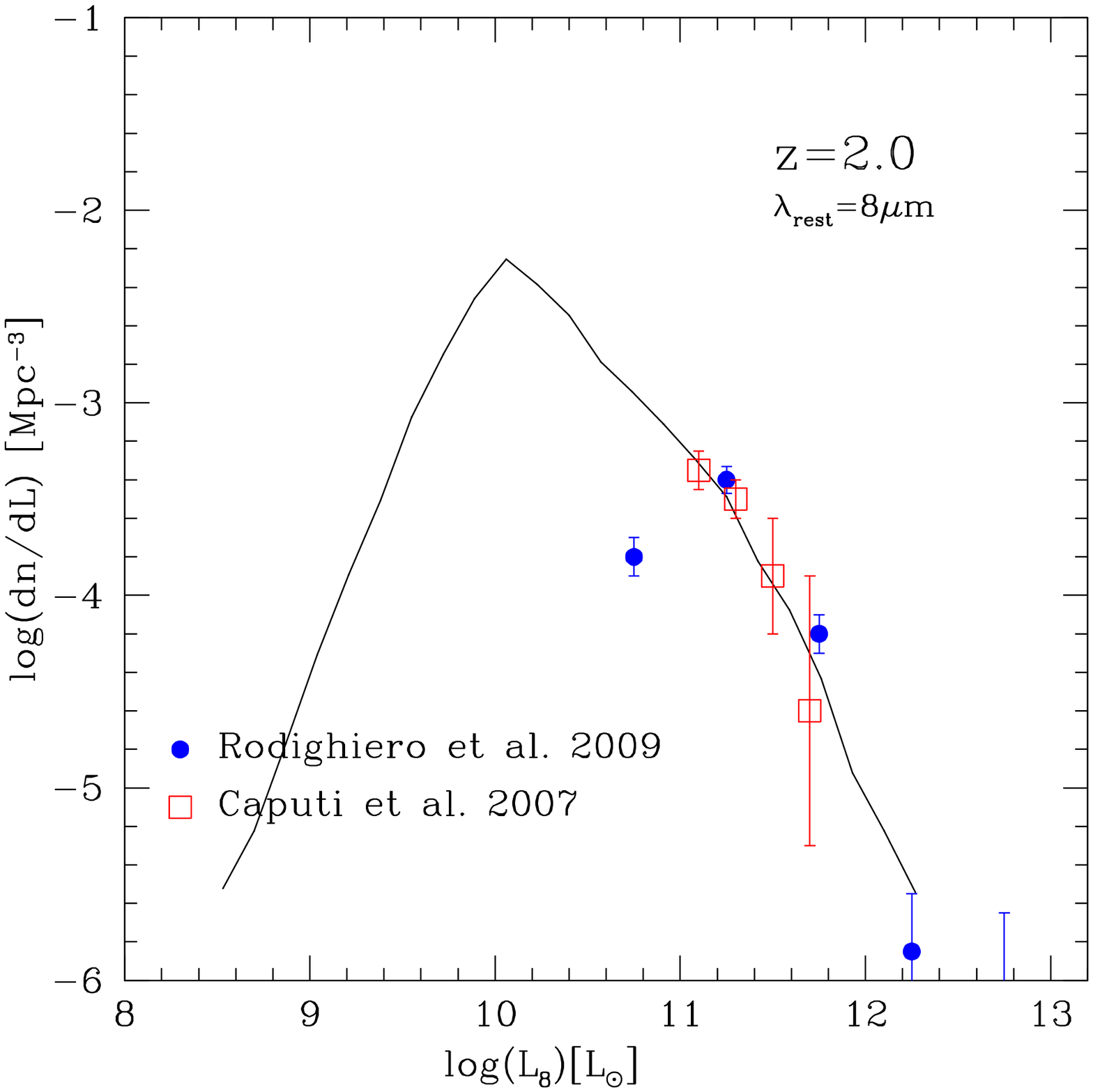}{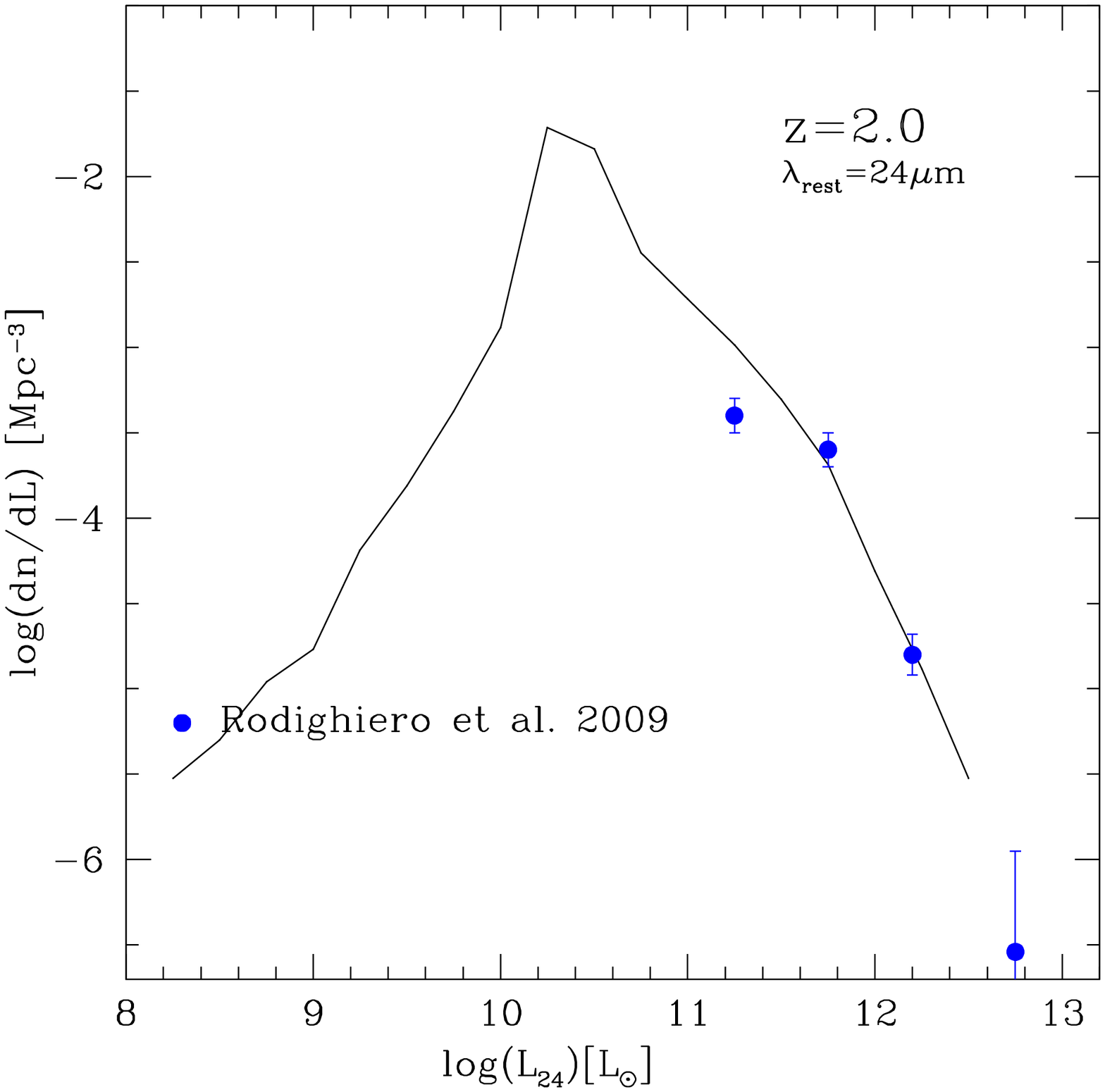}
\caption{Monochromatic rest-frame LFs at 8 and 24\,$\mu$m for $z=2$ simulated galaxies, compared against observational data by \citet{Caputi07} and \citet{Rodighiero09}. The downturn of simulated LFs at low-luminosity end is due to the resolution limit of this simulation with a comoving box size of 10$h^{-1}$Mpc. }
\label{fig:lf_rest}
\end{figure}

Figure\,\ref{fig:lf_rest} compares the monochromatic rest-frame luminosity functions at 8 and 24\,$\mu$m of simulated galaxies at $z=2$ against observed data.  The agreement between the two is quite good, except for the last data point at the lowest luminosity, where the observation might be suffering from the flux limit.

%\section{Summary and Future Work}

\vspace{0.2cm}
{\bf Summary.}
We presented the preliminary results of our initial attempt to calculate the IR emission from high-$z$ galaxies in our cosmological hydrodynamic simulations.  The resulting rest-frame luminosity functions at 8\,$\mu$m and 24\,$\mu$m agree very well with the observations, which is very encouraging.  Through more detailed analyses on the relationship between star formation histories, metallicities, and IR luminosities, we hope to gain more insight on the physical properties of IR galaxies observed by the {\it Spitzer}.

\acknowledgements %%% Text of acknowledgements runs on after this command.

We thank Laura Silva for providing us with the GRASIL code and its output. 
This work is supported by the NASA JPL Spitzer Space Telescope grant 
RSA No. 1347463.
It is also supported in part by the National Aeronautics and Space
Administration under Grant/Cooperative Agreement No. NNX08AE57A issued
by the Nevada NASA EPSCoR program, the President's Infrastructure
Award at UNLV, and by the NSF through TeraGrid resources provided by 
the Texas Advanced Computing Center.  Some simulations were also performed 
at the UNLV Cosmology Cluster.

%%% THE BIBLIOGRAPHY
%%%
%%% CONSULT SECTION 3 OF "INSTRUCTIONS FOR AUTHORS" FOR HOW TO USE NATBIB.
%%% AUTHORS ARE ENCOURAGED TO USE EITHER THE "THEBIBLIOGRAPY" ENVIRONMENT
%%% BY UNCOMMENTING (DELETING THE "%" SYMBOL) THE COMMANDS BELOW, OR BY
%%% USING THE BIBTEX ENVIRONMENT. TO FIND OUT WHICH IS APPLICABLE TO YOUR
%%% CONTRIBUTION, CONSULT THE VOLUME EDITORS FOR YOUR PROCEEDINGS.
%%%

%\bibliographystyle{apj}
%\bibliography{nagamine_k}

\begin{thebibliography}
\expandafter\ifx\csname natexlab\endcsname\relax\def\natexlab#1{#1}\fi

\bibitem[{{Caputi} {et~al.}(2007){Caputi}, {Lagache}, {Yan}, {Dole},
  {et~al.}}]{Caputi07}
{Caputi}, K.~I., {Lagache}, G., {Yan}, L., {Dole}, H., {et~al.} 2007, \apj,
  660, 97

\bibitem[{{Choi} \& {Nagamine}(2010)}]{Choi10}
{Choi}, J. \& {Nagamine}, K. 2010, arXiv:1001.3525

\bibitem[{{Choi} \& {Nagamine}(2009{\natexlab{a}})}]{Choi09a}
{Choi}, J.-H. \& {Nagamine}, K. 2009{\natexlab{a}}, MNRAS, in press, arXiv:0909.5425

\bibitem[{{Choi} \& {Nagamine}(2009{\natexlab{b}})}]{Choi09c}
---. 2009{\natexlab{b}}, \mnras, 393, 1595

\bibitem[{{Kauffmann} \& {Haehnelt}(2000)}]{Kau00}
{Kauffmann}, G. \& {Haehnelt}, M. 2000, \mnras, 311, 576

\bibitem[{{Lacey} {et~al.}(2010){Lacey}, {Baugh}, {Frenk}, {Benson},
  {et~al.}}]{Lacey10}
{Lacey}, C.~G., {Baugh}, C.~M., {Frenk}, C.~S., {Benson}, A.~J., {et~al.} 2010,
  \mnras, 443

\bibitem[{{Lacey} {et~al.}(2007){Lacey}, {Baugh}, {Frenk}, {Silva},
  {et~al.}}]{Lacey07}
{Lacey}, C.~G., {Baugh}, C.~M., {Frenk}, C.~S., {Silva}, L., {et~al.} 2007,
  ArXiv e-prints, 704

\bibitem[{Nagamine {et~al.}(2000)Nagamine, Cen, \& Ostriker}]{Nag00}
Nagamine, K., Cen, R., \& Ostriker, J.~P. 2000, ApJ, 541, 25

\bibitem[{Nagamine {et~al.}(2001)Nagamine, Fukugita, Cen, \& Ostriker}]{Nag01b}
Nagamine, K., Fukugita, M., Cen, R., \& Ostriker, J.~P. 2001, ApJ, 558, 497

\bibitem[{{Rodighiero} {et~al.}(2009){Rodighiero}, {Vaccari}, {Franceschini},
  {Tresse}, {et~al.}}]{Rodighiero09}
{Rodighiero}, G., {Vaccari}, M., {Franceschini}, A., {Tresse}, L., {et~al.}
  2009, arXiv:0910.5649

\bibitem[{Salpeter(1955)}]{Salpeter55}
Salpeter, E.~E. 1955, ApJ, 121, 161

\bibitem[{{Silva} {et~al.}(1998){Silva}, {Granato}, {Bressan}, \&
  {Danese}}]{Silva98}
{Silva}, L., {Granato}, G.~L., {Bressan}, A., \& {Danese}, L. 1998, \apj, 509,
  103

\bibitem[{{Springel}(2005)}]{Springel05e}
{Springel}, V. 2005, \mnras, 364, 1105

\bibitem[{{Springel} \& {Hernquist}(2003)}]{Springel03b}
{Springel}, V. \& {Hernquist}, L. 2003, \mnras, 339, 289

\end{thebibliography}

%\begin{thebibliography}{}
%\bibitem[]{}
%\bibitem[]{}
%\end{thebibliography}

\end{document}